# Surface Enhance Raman Scattering Investigation of Uranium Hydride and Uranium Oxide.

*Martin S. Piltch, Perry C. Gray, and Michael Manley*

## Introduction

This report documents our work on the use of Surface Enhanced Raman Scattering (SERS) as a diagnostic tool to identify and quantify surface layers of oxides and hydrides. First, we review the novel technique, which we are investigating. Our contribution to the field is that we are extending the SERS process to materials that have little or no intrinsic electrical conductivity. This permits the technique to be used on layers of oxides, hydrides and nitrides that have appeared as corrosion products.

The common understanding of the SERS process [1] is based upon the contribution of a high electrical-conductivity substrate or an admixture of silver or gold nanoparticles to enhance the normal Raman signature of a molecular species. Under ordinary circumstances, it is the large electrical conductivity of a material (gold or silver) or an intimately connected substrate, which is responsible for the enhancement of the SERS signals in comparison to the normal Raman effect.

The present approach does not rely on a high electrical conductivity substrate or admixture to provide the SERS enhancement of the normal Raman effect. The novelty of our method is to create an artificial electron distribution on the surface of a normally low-electrical-conductivity material. This distribution is generated by illuminating the metal/oxide/hydride surface with ultraviolet light whose photon energy is above the work function of the material. Since there is no admixed material present, there is no possible contamination of the sample surface. The ultraviolet illumination releases photoelectrons. This constitutes an artificial electrical conductivity as long as the illumination is present. It is this conductivity, in the form of surface polariton plasmons in high conductivity materials such as gold and silver, that leads to the many orders of magnitude enhancement of the normal Raman scattered signal from the oxides and hydrides on these metallic surfaces.

The SERS process is a "parametric" process, that is, a parameter, the virtual photon that is absorbed and re-emitted with energy that is reduced by the molecular vibration energy, is being varied periodically to enhance the Raman signal. When the vector sum of all the wave vectors ($\mathbf{k}=2\pi/\lambda$) in the process equals zero, the SERS effect manifests itself. Let $\mathbf{k}_p$ be the wave vector of the Raman probe radiation, , $\mathbf{k}_s$ that of the Raman signal and the "idler" wave vector be $\mathbf{k}_i$. When ($\mathbf{k}_p$-$\mathbf{k}_s$-$\mathbf{k}_i$) is zero, the SERS effect takes place. The idler wave relates to the surface polariton plasmon or in our case, the ultraviolet produced electrical conductivity. We assure this happening by producing the ultraviolet radiation-induced electrical conductivity on the surface in a spatially periodic manner. This periodicity simulates the presence of the surface polariton plasmon in the conventional SERS process. Illuminating the surface to be investigated with an ultraviolet interference pattern or diffraction pattern produced by a multiple slit diffracting obstacle experimentally produces the periodic electrical conductivity. The periodicity of illumination is controlled by choosing the characteristics of the diffraction obstacle. We chose a laser (Coherent MBD266) that fulfills the requirements of short ultraviolet wavelength and coherence length sufficient to produce an extended diffraction pattern on the non-metallic surface. It makes use of state–of-the-art semiconductor diode pumped neodymium doped lithium vanadate laser material. The 1.06$\mu$m infrared output of this crystal is internally frequency-doubled to a wavelength of 532nm. This radiation is then applied to a second frequency doubler (BBO) that converts the 532nm green radiation to usable ultraviolet at 266 nm. This device produces 200mW of continuous, single longitudinal mode power. The latter property gives it a coherence length suitable for projecting a double-slit (1-2 centimeter spacing) interference pattern on an oxidized surface. As can be seen from the TABLE 1, this wavelength, having photon energy of 4.66eV is sufficiently short to produce photoelectrons in all relevant materials.

| Element | WF(ev) | WF(nm) | WF Oxide(ev) | WF Oxide(nm) |
|---|---|---|---|---|
| U | 3.45 | 358 | 3.1 | 400 |
| Fe | 4.5 | 276 | | |
| Ni | 4.41 | 281 | 1.82[6] | 676[6] |
| Cu | 4.4 | 282 | 4.70[8] | 264[8] |
| Ir | 4.87 | 255 | 4.23[7] | 293[7] |
| Ag | 4.4 | 282 | | |
| Nb | 4.2 | 295 | | |
| Ti | 4.0 | 310 | | |

**TABLE 1 Work function of oxides and hydrides**

The laser that we have chosen for the Raman probe is a green 532nm device based upon the neodymium ion in various hosts.  This short wavelength is ideal for a Raman probe because the Raman intensity varies directly as the fourth power of the laser frequency.  The laser needs to have continuous wave output as well as a linewidth, which is sufficiently narrow so that the Raman signature of a typical oxide or hydride molecule (approximately100 to 400 wavenumbers) can be resolved.

We developed a simple, fiber optics coupled Raman spectrometer of sufficient sensitivity and resolution to observe the normal Raman scattering signal from molecules that are considered standards.  We observed the Raman signals from standards chosen with Raman shifts in the range of those expected from metal oxides and hydrides of interest.  We then observed normal Raman scattering from the oxides and hydrides of interest.  Using slurries of oxides mixed with colloidal gold particles we observed the signal augmentation due to the Surface Enhanced Raman Scattering parametric process. We then measured the enhancement due to application of oxide slurries to high conductivity gold plated surfaces.  Finally, we used the ultraviolet laser to illuminate the oxide surface to create the artificial photoelectron periodic surface conductivity, which produce the surface polariton plasmon density that is responsible for SERS.

**Experimental activities**

Our first Raman spectrometer used a 1/4 meter focal length monochromator (Chromex model 250SM). This device had three computer selectable diffraction gratings that covered the visible through near-infrared regions of the spectrum. The instrument is capable of computer controlled wavelength scans throughout these regions. It is equipped with a type 928 photomultiplier tube at its output slit. It has a resolution of at least 50,000 and very low scattered inter-order light, making it ideal as a Raman instrument. As a light source, we used two lasers, a standard helium-neon laser having a 632.8 nm wavelength and a small, frequency doubled, diode-pumped Nd:YAG laser at 532nm. The latter has the advantage of the increase in Raman sensitivity that varies as the fourth power of the wavelength ratio. The laser beam, prior to impinging on the sample to be investigated, was chopped by a Signal Recovery model 651 optical chopper. The Raman signal, after detection by the photomultiplier, was amplified by a Signal Recovery type 5183 low-noise preamplifier and finally processed by a Signal Recovery model 7225 Digital Phase-Sensitive Detector. Using this apparatus, we performed spectral scans of many compounds that are widely accepted as Raman standards. These included 4- acetamidophenol which has a Raman signature that includes lines at 651 and 857 cm$^{-1}$.

We then prepared aqueous slurries of $Y_2O_3$, $Nd_2O_3$ and $Pr_6O_{11}$ on the surface of high conductivity gold plated substrates. The Yttrium oxide is of particular interest in that it has a strong Raman shift at 380cm$^{-1}$. The Raman shifts for various Uranium oxides appear in the recent literature. They are presented in the table below.

| U Oxide Species | Raman Shift (cm-1) | Band Assignment |
|---|---|---|
| $UO_2$ | 450 | U-O stretch |
| $UO_3$ | 767, 838 | O-U-O-U, U-O stretch |
| $U_3O_8$ | 343, 412, 740, 811 | U-O, U-O, O-U-O-U, U-O stretch |

**TABLE 2) Vibrational modes of uranium oxides [2,3].**

At this stage, we changed Raman spectrometers from the conventional Chromex monochrometer to a modern, fiber optics coupled, CCD array device manufactured by Ocean Optics, Inc. This was to facilitate rapid measurement of Raman spectra. The CCD array detector along with its integrating software also had better noise characteristics than the Chromex photomultiplier. The software permitted a Raman spectrum to be acquired in 30 seconds as compared to the ten minutes required for the conventional system. We used fiber optics for acquiring the Raman signals. Holographic, band-rejection filters provided sufficient attenuation to insure that probe beam scattered light did not compromise the resolution of the system by overloading the detector. We generated the appropriate periodic illumination using a double slit interference pattern. The optical double-slits for producing the periodic illumination pattern were specially laser-cut by Lenox Lasers Inc.

We designed and fabricated a sample cell to permit in-situ hydriding of various materials. UV grade sapphire windows were mounted to a gas-tight cell to permit sample preparation and illumination by both the uv radiation as well as by the Raman probe radiation. A mount was fabricated for the input and output fibers with appropriate lenses to acquire the Raman scattered signal and transmit it to the CCD spectrometer. Located between the Raman signal input fiber and the relay fiber that conveys the signal to the spectrometer is a pair of collimating lenses. These lenses produce nearly parallel light from the highly diverging light from the input fiber (numerical aperture of 0.27) to a region where the holographic "notch", band reject filter is located. This filter is used to attenuate the unshifted 532 nm laser energy that is specularly reflected from the target prior to the beam going into the spectrometer for analysis and wavelength measurement. This augments the wavelength discrimination ability of the spectrometer. It offers an attenuation of approximately $10^4$ at the Raman probe wavelength. When coupled with the spectrometer's resolution of approximately $10^4$ the combination provides more than adequate discrimination to permit weak Raman signals to be detected that are within a few hundred wavenumbers of the probe. By proper angular tuning, the filter permitted observation of Raman features less than 100 wavenumbers removed from the excitation wavelength.

Using the above apparatus, we made conventional Raman measurements of $CaF_2$ and $Y_2O_3$ having Raman shifts of 380 and 322 wavenumbers, respectively. The output of the detector showed numerous Raman peaks, which were quite reproducible upon changing fibers. In all measurement cases we observed a series of Stokes shifted peaks that are attributable to the $CaF_2$ target. The noise level was low enough such that the $CaF_2$ peaks were displayed with a signal to noise ratio of at least 10:1. Thus, we developed confidence that surfaced enhanced Raman signals should appear with S/N rations of at least $10^6$.

Using the CCD spectrometer, we made conventional backscatter measurements of the Raman active modes of boron nitride (BN) and of sodium bromate ($NaBrO_3$). Both of these inorganics have strong Raman modes that were easily detectable with our fiber optical coupled apparatus. We performed surface enhanced Raman scattering using an aqueous solution of sodium bromate mixed with 5 micron diameter gold powder applied to a glass slide. The solution is permitted to evaporate leaving an inhomogeneous mixture of gold particles and sodium bromate crust on the surface of the slide. This arrangement displayed a surface enhancement of the signal from the normal Raman active modes. Surface enhanced Raman scattering was also observed using gold plated 304 stainless steel targets upon which a similar solution of sodium bromate was evaporated. We also observed surface enhanced Raman scattering using a mixture of sodium bromate and gold powder in a standard spectroscopic grade quartz cuvette. This produced the largest enhancement of the Raman signal.

The above measurements of SERS effects have as their basis the fact that a high electrical conductivity material such as gold can provide the required surface polariton plasmons that are responsible for the signal enhancement associated with SERS. Using the three methods described above, we have sought to obtain as intimate a mixture as possible of the Raman active material and the high conductivity metallic powder to enable the SERS effect.

We made extensive measurements of the strongly Raman-active mode[4] of zinc oxide

(ZnO) at 420 cm$^{-1}$. These measurements were made first using a quartz cuvette filled with the ZnO powder, illuminated by the 40mW Nd:YAG laser focused with a lens of 100mm focal length. The Raman signal was acquired by the usual multimode fiber in close proximity to the ZnO target. Next, a 100 $\mu$l microcapilliary tube was filled with the ZnO powder, illuminated with the laser and the measurements successfully repeated. The ZnO powder was mixed with distilled water to make a thick slurry suspension. It was applied to a metal disc with a nickel surface. The water was allowed to evaporate and the ZnO adhered to the surface. The Raman measurement was successfully repeated. Surface enhanced Raman measurements were made by mixing 1:1 by volume, the ZnO along with 5$\mu$m diameter gold powder. This mixture was first illuminated in the quartz cuvette and then by producing an air-dried slurry. In each case, a surface enhanced Raman signal was observed due to the intimate mixture of the insulating ZnO with the high electrical conductivity gold particles.

A convenient vacuum-compatible sample holder was fabricated that aligns small sample discs near the window through which the probe laser beam propagates. This device consists of a nominal 1 3/4 inch high vacuum "T" fitting equipped with ultraviolet grade sapphire windows and a high vacuum, bakeable valve. The disc samples are held and located using slots that are milled in an aluminum insert, which rests on the bottom of the "T". With the vacuum windows in place, similar measurements were made on ZnO on metallic disc targets as described above. Measurements were repeated with evaporated solutions of sodium bromate as well as sodium bromate/gold particle slurries as the Raman active material to confirm that the system was ready to perform measurements on hydrides.

## Measurements on uranium compounds

The vacuum-compatible sample holders were loaded with four uranium compounds, $UO_2$, $UO_3$, $UO_4$ and $U_3O_8$. Each one of the crystalline materials was adhered to a 304 stainless steel disc that was held in the sample holder. These materials were loaded into the sample holders inside a helium atmosphere glovebox. Since none of the samples were prone to further oxidation, the sample holders were brought into the local air environment upon removal from the glovebox. The optical windows were then bolted on to their mating flanges with crushable copper gaskets to assure no escape of material or further ingress of the atmosphere.

We illuminated uranium compounds with the spectrometer. The Raman signatures of these materials [5] are well documented. We easily observed the 400cm$^{-1}$ to 800cm$^{-1}$ signatures from the $UO_2$ and $UO_3$ and proceeded to the next set of samples.

We worked on the optical system to observe the expected Raman signature form uranium hydride at 85 cm$^{-1}$. This small Raman shift is due to the fact that the $UH_3$ molecule is dominated in its vibrational modes by the massive uranium atom. The technique that we are using to discriminate against the Raman excitation wavelength is two holographic filters in series. We will use another airtight target holder to perform Raman experiments on $UH_3$ and then continue to the measurement of the ingrowth rate of the hydride using ultraviolet radiation to generate photoelectrons.

# Initial Enhanced Raman Measurements on Uranium Hydride

As an initial attempt at obtaining a Raman signature from uranium hydride, direct measurements were made on $UH_3$ powder; in which the oxides were also expect to be present. Two samples holders were assembled from stainless steel high vacuum flanges with sapphire optical windows. These two assemblies were loaded with small amounts of $UH_3$ powder and $UH_3$ powder mixed with fine gold powder. The excitation source used for the Raman measurements was a Spectra Physics Millennium V diode pumped frequency doubled NdYAG laser operated at power levels between 0.2 and 0.5 Watts. Light was picked up by direct scatter illumination of an optical fiber, which directed the signal through holographic notch filters to eliminate the excitation light and then into our fiber coupled imaging spectrometer (Ocean Optics Inc., HC2000). Because the Raman shift for the hydride is small, the angle of the notch filter was adjusted to position the excitation wavelength (532 nm) near the long wavelength edged of the notch, making maximum signal from the hydride Raman signal available. The excitation source was held a 0.3 Watts (maximum from this laser is 5 Watts). The integration time for the signal was held constant at 5 seconds. Figure 1 shows the results of comparative runs between the two samples,

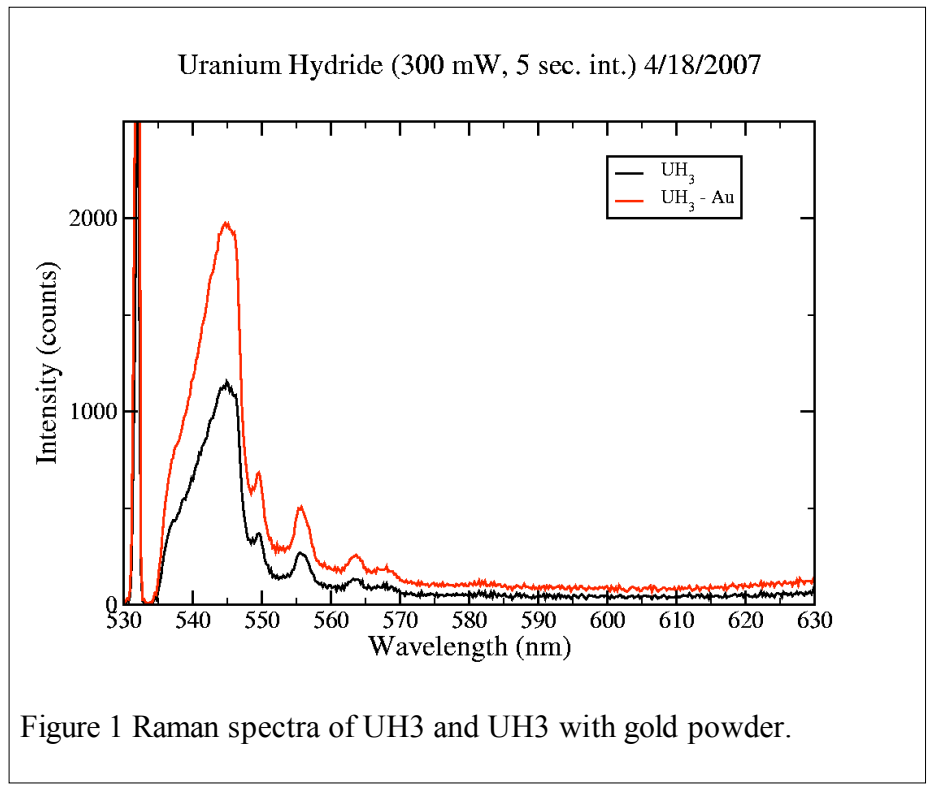

Figure 1 Raman spectra of UH3 and UH3 with gold powder.

The major peaks seen are due to the oxides formed from exposure to air, with the main $UO_3$ peak a broad feature at 544.5. Even though the filter was adjusted to maximize the hydride Raman peak, it is expected at 534.5 nm just 2.5 nm from the excitation source at 532 nm (seen as a sharp spike). The hydride is seen as an unresolved feature, the shoulder on the main oxide peak. The addition of gold powder to the sample results in a marked enhancement of the Raman signal indicating a SERS effect.

## Conclusions

We have demonstrated the SERS process on various non-conductive materials including the oxides and hydride of uranium. The long range goal of the project, that of using ultraviolet illumination to produce a plasmon-like conductivity was investigated but was not unequivocally achieved. We will concentrate on this activity when the project resumes in the coming year. In the near future we will also try several different approaches to increasing the resolution of the detection system to resolve the hydride peak. The imaging spectrometer used in these measurements, while compact, is not optimized for spectral resolution.